\documentstyle[12pt]{ioplppt}
\begin{document}
\jl{6}


\def\be{\begin{equation}}
\def\ee{\end{equation}}
\def\bea{\begin{eqnarray}}
\def\eea{\end{eqnarray}}
\def\l{\label}
\def\ct{\cite}
\def\r{\ref}
\def\Th{\Theta}
\def\sig{\sigma}
\def\om{\omega}
\def\udot{\dot{u}}
\newcommand{\sfrac}[2]{{\textstyle{#1\over#2}}}
\def\case#1/#2{\textstyle\frac{#1}{#2}}


\title{Integrability of irrotational silent cosmological
models}

\author{Henk van Elst\dag\ $^1$,
Claes Uggla\ddag\S\ $^2$,
William M Lesame\P\ $^3$,
George F R Ellis$\ast$\ $^4$
and Roy Maartens$\sharp$\ $^5$}

\address{\dag\ Astronomy Unit, Queen Mary \& Westfield College,
University of London, London E1 4NS, Britain}
\address{\ddag\ Department of Physics, Stockholm University,
Box 6730, S--113 85 Stockholm, Sweden}
\address{\S\ Department of Physics, Lule\aa\ University of
Technology, S--951 87 Lule\aa, Sweden}
\address{\P\ Department of Applied Mathematics, University of
Fort Hare, Private Bag X1314, Alice 5700, South Africa}
\address{$\ast$\ Department of Applied Mathematics, University of
Cape Town, Cape Town 7700, South Africa}
\address{$\sharp$\ School of Mathematical Studies, Portsmouth
University, Portsmouth PO1 2EG, Britain}

\begin{abstract}

We revisit the issue of integrability conditions for the
irrotational silent cosmological models. We formulate the problem
both in $1+3$ covariant and $1+3$ orthonormal frame notation, and
show there exists a series of constraint equations that need to be
satisfied. These conditions hold identically for FLRW--linearised
silent models, but not in the general exact non--linear case.
Thus there is a linearisation instability, and it is highly unlikely
that there is a large class of silent models.
We conjecture that there are no spatially inhomogeneous
solutions with Weyl curvature of Petrov
type I, and indicate further issues that await clarification.

\end{abstract}

\pacs{0420, 9880H, 9880D, 0420J}

\vfill
{\footnotesize
\noindent
$^1$h.van.elst@maths.qmw.ac.uk\\
$^2$uggla@vanosf.physto.se\\
$^3$lesame@ufhcc.ufh.ac.za\\
$^4$ellis@maths.uct.ac.za \\
$^5$maartens@sms.port.ac.uk
}
\newpage

\section{Introduction}

The idea behind the silent cosmological models, introduced and
discussed by Matarrese et al (\,see Refs. \ct{matetal94},
\ct{bruetal95b} and \ct{bruetal95}\,), is the following. Starting
from any consistent initial configuration for barotropic perfect
fluid spacetime geometries $\left(\,{\cal M}, \,{\bf g}, \,{\bf u}
\,\right)$, there exist two physically different phenomena that
convey information between adjacent worldlines within the preferred
timelike reference congruence ${\bf u}$,
which is identified with the average
4-velocity of the fluid matter source. These are sound waves and
gravitational waves.  Mathematically, in the $1+3$ covariant
representation of such models (\,see Refs. \ct{ehl61}, \ct{ell71},
\ct{ell73} and \ct{ell96}\,), 
these dynamical interactions are represented by
the spatial derivative source terms in the evolution equations. A
careful look at the perfect fluid time derivative equations,
obtained from the Ricci identities for ${\bf u}$ and the Bianchi
identities, reveals that these terms are comprised of either the
spatial derivatives of the fluid acceleration, or the spatial
rotation terms (often also called the `curls') of both the
electric and magnetic parts of the Weyl curvature,     
$E_{\mu\nu}$ and $H_{\mu\nu}$. Given the barotropic equation of
state assumption, the fluid acceleration, and consequently the
spatial 3--gradient of the fluid pressure, are linked to the
spatial 3--gradient of the fluid's total energy density. Pressure
3--gradients and their spatial variations generate propagating
sound waves within the fluid, while non--zero values of the spatial
rotation of either of the Weyl curvature variables generically
induce temporal changes in the other, usually interpreted as
propagating gravitational waves.

Both from a mathematical and a physical point of view, it is of
interest to investigate the case in which these spatial derivative
terms vanish, so that the resultant $1+3$ covariant evolution
equations reduce to a coupled set of first--order ordinary
differential equations, in terms of the comoving time derivative
along the fluid flow lines. In this case, provided the constraint
equations (involving only orthogonal spatial derivatives) are
satisfied on an initial surface and remain satisfied, the
subsequent evolution along individual fluid worldlines within ${\bf
u}$ is decoupled,   
i.e., the covariant time derivatives decouple from the orthogonally
projected spatial ones.  This technically describes what is called
the silent assumption for cosmological models; it expresses the
physical idea of an absence of any form of waves or of
gravitational induction (hence, any form of causal communication)
propagating between the worldlines of neighbouring fluid
elements. Thus, there is no exchange of new information (i.e.,
information not already coded in the initial data) between the
fluid elements.

To realise it, Matarrese et al assumed the fluid matter source to
be irrotational dust, which generates a spacetime geometry of
purely electric Weyl curvature, i.e., 
the magnetic part of the Weyl curvature vanishes\footnote{Here and
in the rest of this paper, we use the standard $1+3$ covariant
formalism, thoroughly reviewed in Ref. \ct{ell71} (\,see also
Refs. \ct{ell73}, \ct{ell96},
\ct{ellbru89}, \ct{matetal94}, \ct{bruetal95b}
and \ct{bruetal95}\,). Sign and index conventions are employed
according to Ref. \ct{elsugg96}.}:
\be
\l{sil}
\om^{\mu} = 0 \,, \hspace{1cm} p = 0 \Rightarrow \udot^{\mu} = 0\,,
\hspace{1cm} H_{\mu\nu} = 0 \ .
\ee
Well--known exact solutions of the Einstein field equations fall
into this category. Spatially inhomogeneous representatives are the
dust spacetime geometries given by Szekeres \ct{sze75} (\,discussed
in a nice geometrical formulation by Goode and Wainwright
\ct{goowai82}\,), as well as              
Ellis' dust subclass of LRS class II spacetime geometries, which
includes the Lemaitre--Tolman--Bondi model and the
orthogonally spatially homogeneous (OSH) Kantowski--Sachs model
(\,see Refs.  \ct{ell67} and \ct{elsell96}\,). In both examples,
the Weyl curvature tensor is of algebraic Petrov type D (\,see,
e.g., Ref.  \ct{ksmh80}\,). A Petrov type I example of a silent
model is provided by the OSH Bianchi Type--I dust solutions (\,see,
e.g., Ref. \ct{ellmac69}\,).

The question is how many other such silent solutions there are.
This depends on the consistency of the constraint equations with
the time evolution equations. A previous paper \ct{lesetal95}
claimining that these equations are generically consistent with
each other is, regrettably, wrong, as        
indicated in \ct{maa96} and demonstrated by Bonilla et al
\ct{spa96}. However, the latter paper did not determine the set of
consistent solutions. We mount a systematic attack on that question
here.

In the following, we first give the dynamical equations defining
irrotational dust silent spacetimes in $1+3$ covariant
form; this is done in section \r{ch5silcov}. We derive a   
covariant integrability condition, and show that it is
satisfied at linear level, i.e., 
for those silent models which are                            
linearised inhomogeneous perturbations of FLRW dust universes.
However the condition is non--trivial in the non--linear case. 
Thus {\em there is a linearisation instability in the silent models.}
In section \r{ch5silonf} the silent configurations are then
formulated in $1+3$ orthonormal frame (ONF) terminology. By the
assumption of vanishing magnetic Weyl curvature a {\em new\/}
constraint is generated. The consistency of this constraint with
the remaining dynamical equations is investigated in section
\r{ch5silcons}. We show that a sequence of non--trivial
differential and algebraic conditions results from repeated
propagation of the constraints along the integral curves of ${\bf
u}$. {\em These conditions are not identically
satisfied in general irrotational silent spacetimes\/}, 
throwing into question the further physical analysis of silent models,
and we conjecture, in particular, that there are no consistent
spatially inhomogeneous solutions with a Weyl curvature tensor of
algebraic Petrov type I.  Finally, the results obtained are
discussed in section \r{ch5silres}.

\section{$1+3$ covariant formulation}
\l{ch5silcov}

The silent conditions Eq. (\r{sil}) describe irrotational and
pressure--free fluid matter sources inducing purely electric Weyl
curvature. The $1+3$ covariant dynamical equations then take the
form\\

\noindent
{\em Time derivative equations}
\bea
\l{csilthdot}
\dot{\Th} & = & - \,\sfrac{1}{3}\,\Th^{2} - 2\,\sig^{2}
- \sfrac{1}{2}\,\mu \\
\l{csilsigdot}
h^{\mu}\!_{\rho}\,h^{\nu}\!_{\sig}\,(\dot{\sig}^{\rho\sig})
& = & - \,\sfrac{2}{3}\,\Th\,\sig^{\mu\nu} - \sig^{\mu}\!_{\rho}\,
\sig^{\nu\rho} - E^{\mu\nu} + \sfrac{2}{3}\,\sig^{2}\,
h^{\mu\nu} \\
\l{csiledot}
h^{\mu}\!_{\rho}\,h^{\nu}\!_{\sig}\,(\dot{E}^{\rho\sig})
& = & - \,\sfrac{1}{2}\,\mu\,\sig^{\mu\nu} - \Th\,E^{\mu\nu}
+ 3\,\sig^{(\mu}\!_{\rho}\,E^{\nu)\rho} - \sig^{\rho}\!_{\sig}
\,E^{\sig}\!_{\rho}\,h^{\mu\nu} \\
\l{csilmudot}
\dot{\mu} & = & - \,\Th\,\mu \ .
\eea

\noindent
{\em Constraint equations}
\bea
\l{csildivsig}
0 & = & h^{\mu}\!_{\rho}\,h^{\nu}\!_{\sig}\,(\nabla_{\nu}
\sig^{\rho\sig}) - \sfrac{2}{3}\,Z^{\mu} \ := \ (C_{1})^{\mu} \\
\l{csildive}
0 & = & h^{\mu}\!_{\rho}\,h^{\nu}\!_{\sig}\,(\nabla_{\nu}
E^{\rho\sig}) - \sfrac{1}{3}\,X^{\mu} \ := \ (C_{2})^{\mu} \\
\l{csildivh}
0 & = & \eta^{\mu\nu\rho\sig}\,\sig_{\nu\tau}\,
E^{\tau}\!_{\rho}\,u_{\sig} \ := \ (C_{3})^{\mu} \\
\l{csilhconstr}
0 & = & K_{\mu\nu} \ := \ h^{\rho}\!_{(\mu}\,h^{\sig}\!_{\nu)}
\,\eta_{\rho\tau\kappa\lambda}\,(\nabla^{\tau}
\sig^{\kappa}\!_{\sig})\,u^{\lambda}
\ = \ [\,\nabla\times\sig\,]_{\mu\nu} \ := \ (C_{4})_{\mu\nu}
\\
\l{csilhdot}
0 & = & I_{\mu\nu} \ := \ h^{\rho}\!_{(\mu}\,h^{\sig}\!_{\nu)}
\,\eta_{\rho\tau\kappa\lambda}\,(\nabla^{\tau}
E^{\kappa}\!_{\sig})\,u^{\lambda}
\ = \ [\,\nabla\times E\,]_{\mu\nu} \ := \ (C_{5})_{\mu\nu} \ .
\eea

\noindent
{\em Gau\ss\ equations}
\bea
{}^{3}\!S_{\mu\nu} & = & E_{\mu\nu} - \sfrac{1}{3}\,\Th
\sig_{\mu\nu} + \sig_{\mu\rho}\,\sig^{\rho}\!_{\nu}
- \sfrac{2}{3}\,\sig^{2}\,h_{\mu\nu} \\
{}^{3}\!R & = & 2\,\mu - \sfrac{2}{3}\,\Th^{2} + 2\,\sig^{2} \ .
\eea
The constraints in this setting have the following implications.
The spatial divergence of the fluid rate of shear $\sig_{\mu\nu}$
is related to the spatial 3--gradient of the fluid rate of
expansion, $Z_{\mu} := h^{\nu}\!_{\mu}\nabla_{\nu}\Th$,
Eq. (\r{csildivsig}), and, analogously, the spatial divergence of
the electric part of the Weyl curvature $E_{\mu\nu}$ to the spatial
3--gradient of the fluid's total energy density, $X_{\mu} :=
h^{\nu}\!_{\mu}\nabla_{\nu}\mu$, Eq. (\r{csildive}).  Both the
fluid rate of shear tensor and the electric part of the Weyl
curvature tensor share a common eigenframe, a property expressed by
Eq. (\r{csildivh}). This result was originally obtained by Barnes
and Rowlingson \ct{barrow89}.  Additionally, the constraints
(\r{csilhconstr}) and (\r{csilhdot}) express the condition that in
silent configurations as defined by Eq. (\r{sil}), these tensors
need to have zero spatial rotation.

Maartens \ct{maa96} has shown that the constraints for generic 
irrotational dust spacetime geometries (with non--zero
$H_{\mu\nu}$) are consistent with each other and preserved along
the integral curves of ${\bf u}$, if they are satisfied at an
initial instant. This result has been extended by van Elst
\ct{hve96} to generic barotropic perfect fluid spacetimes (with
non--zero $H_{\mu\nu}$ and $E_{\mu\nu}$). Equations
(\r{csildivsig}) -- (\r{csilhconstr}) are just reduced forms of the
constraints underlying general barotropic perfect fluid spacetime
geometries. The character of Eq. (\r{csilhdot}), however, is
slightly different. It is a {\em new\/} constraint arising as a
consequence of imposing the silent conditions, Eq. (\r{sil}).
Namely, the vanishing of the magnetic part of the Weyl curvature,
$H_{\mu\nu}$, results in the reduction of the general evolution
equation for $H_{\mu\nu}$ to a constraint, Eq. (\r{csilhdot}).
This is an example of the conversion process where the imposed
assumption that a dynamical variable remains zero throughout the
time evolution results in a further integrability condition,
additional to those generically obtained from the general $1+3$
covariant dynamical equations. Consistency of Eq. (\r{csilhdot})
with the remaining set of equations demands that
\be
\l{csilcurledot1}
0 = h^{\mu}\!_{\rho}\,h^{\nu}\!_{\sig}\,(\dot{I}^{\rho\sig})
= h^{\mu}\!_{\rho}\,h^{\nu}\!_{\sig}\,(\dot{C}_{5})^{\rho\sig}
\ee
holds throughout. Using the methods and identities developed in
Ref. \ct{maa96}, this is equivalent to
\bea
\l{csilcurledot2}
0 & = & - \,\sfrac{4}{3}\,\Theta\,(C_{5})^{\mu\nu}
- \sfrac{3}{2}\,E^{(\mu}\!_{\rho}\,\eta^{\nu)\rho\sig\tau}\,
(C_{1})_{\sig}\,u_{\tau} - \sfrac{3}{2}\,\sig^{(\mu}\!_{\rho}\,
\eta^{\nu)\rho\sig\tau}\,(C_{2})_{\sig}\,u_{\tau} \nonumber \\
& &   - \ \sfrac{1}{2}\,\mu\,(C_{4})^{\mu\nu}
+ \sfrac{3}{2}\,B^{\mu\nu} \ ,
\eea
where
\bea
\l{csilcurledot3}
B^{\mu\nu} & := & h^{(\mu}\!_{\rho}\,h^{\nu)\sig}\,
\eta^{\rho\tau\kappa\lambda}u_\lambda\sig^{\xi}\!_{\tau}\,
\left(\nabla_{\xi}E_{\kappa\sig}\right)
- 3\,h^{(\mu}\!_{\rho}\,h^{\nu)\sig}\,
\eta^{\rho\tau\kappa\lambda}u_\lambda\nabla_{\tau}\left[
\ \sig^{\xi}\!_{(\kappa}\,E_{\sig)\xi}\ \right]
\nonumber \\
& &   - \ \left[\ \sfrac{1}{2}\,\sig^{(\mu}\!_{\rho}\,
X_{\sig} + E^{(\mu}\!_{\rho}\,Z_{\sig}\ \right]\,
\eta^{\nu)\rho\sig\lambda}u_\lambda \ .
\eea
Hence, using Eqs. (\r{csildivsig}) - (\r{csilhdot}),
\be
\l{csilcurledot4}
0 = B^{\mu\nu} \ .
\ee
The $1+3$ covariant condition (\r{csilcurledot4}) may be simplified
and clarified by rewriting Eq. (\r{csilcurledot3}) in terms of the
irreducible parts of the spatial covariant derivatives of
$\sig_{\mu\nu}$ and $E_{\mu\nu}$. This is achieved via the   
complete covariant decomposition of the spatial derivative of any
rank 2 symmetric tracefree 
tensor field $A_{\mu\nu}$ orthogonal to ${\bf u}$, presented in Ref.
\ct{maaetal96}:
\bea
\l{S}
h^{\sig}\!_{\mu}\,h^{\tau}\!_{\nu}\,h^{\kappa}\!_{\rho}\,
\left(\nabla_{\sig}A_{\tau\kappa}\right) & = &\widehat{A}_{\mu\nu\rho}
+ \sfrac{3}{5}\,h_{\mu<\nu}\,h^{\sig}\!_{\rho>}\,
h^{\tau}\!_{\kappa}\,\left(\nabla_{\tau}A^{\kappa}\!_{\sig}\right) 
\nonumber\\
&&- \sfrac{2}{3}\,[\,\nabla\times A\,]^{\sig}\!_{(\nu}\,
\eta_{\rho)\mu\sig\tau}\,u^{\tau} \ ,
\eea
where angle brackets enclosing indices denote the 
spatially projected symmetric tracefree part, and 
$\widehat{A}_{\mu\nu\rho} = \widehat{A}_{<\mu\nu\rho>}$ is the
divergence--free and irrotational part of the spatial derivative of
$A_{\mu\nu}$, known as the `distortion' of $A_{\mu\nu}$.
By Eqs. (\r{csilhconstr}) and (\r{csilhdot}), the spatial rotation
terms in Eq. (\r{S}) vanish for $A_{\mu\nu} = \sig_{\mu\nu}$ and
$A_{\mu\nu} = E_{\mu\nu}$ in irrotational silent models. The
spatial divergence terms are determined by Eqs. (\r{csildivsig})
and (\r{csildive}). Then Eq. (\r{csilcurledot3}) may be reduced to
a form in which all the terms are                             
spatial divergences or distortions of the fluid rate of shear and
the electric part of the Weyl curvature:
\bea
\l{B}
B^{\mu\nu} & = & 
\left[\widehat{\sig}^{\kappa(\mu}\!_{\rho}\,E_{\sig\kappa}
+ \sfrac{2}{3}\,\widehat{E}^{\kappa(\mu}\!_{\rho}\,\sig_{\sig\kappa}
+ \sfrac{8}{45}\,\sig^{(\mu}\!_{\rho}\,X_{\sig}
+ \sfrac{1}{15}\,E^{(\mu}\!_{\rho}\,Z_{\sig}\right]
\eta^{\nu)\rho\sig\tau}\,u_{\tau}\,.
\eea
The form of Eq. (\ref{B}) makes clear the separate roles of the
spatial divergence terms and the totally symmetric tracefree parts
of the derivatives.

Lesame et al \ct{lesetal95} recently reported that condition
(\r{csilcurledot4}) was identically satisfied. However, this is
incorrect \ct{spa96}, and a different conclusion is obtained in
what follows. On the other hand, considering covariant and
gauge--invariant linearised perturbations
of FLRW spacetimes in the sense of Ellis and Bruni
\ct{ellbru89}, Eq. (\r{csilcurledot4}) {\em is\/} identically
satisfied, as all terms occurring in Eq. (\r{csilcurledot3}) are of
second order. Thus, in FLRW--linearised silent models the
constraints evolve consistently. As is seen from our subsequent
analysis, Eq. (\r{csilcurledot4}) is {\em not\/} identically
satisfied in the exact non--linear case. Consequently this leads to
a linearisation instability in irrotational silent cosmological
models. In other words, there would be consistent FLRW--linearised
solutions, which do {\em not\/} correspond to consistent exact
solutions. A constraint similar to Eq. (\r{csilhdot}) arises in the
context of the work by Barnes and Rowlingson \ct{barrow89}, who,
among the conditions (\r{sil}), allowed the fluid pressure to be
{\em non--zero\/} instead (in which case, of course, a
configuration no longer can be called silent). However, in their
work the integrability of that constraint was never established.

This concludes the $1+3$ covariant discussion of the irrotational
silent models. To pursue a detailed constraint analysis, for which
purpose we make use of algebraic computing facilities, it turns out
to be of (relative) computational ease to describe the problem
within the related $1+3$ ONF framework instead. After reformulating
the set of dynamical equations (\r{csilthdot}) -- (\r{csilhdot}) in
terms of $1+3$ ONF variables, the consistency of Eq.
(\r{csilhdot}) with the remaining set is further investigated.

\section{$1+3$ ONF formulation}
\l{ch5silonf}

The ONF approach employs a set of linearly independent 1--form
fields $\{\,\mbox{\boldmath $\omega$}^{a}\,\}$ defined at each
point of the spacetime manifold $\left(\,{\cal M}, \,{\bf g}\,
\right)$ such that the line element can be locally expressed as
\be
\l{add1}
ds^{2} = \eta_{ab}\,\mbox{\boldmath $\omega$}^{a}\,
\mbox{\boldmath $\omega$}^{b} \ ,
\ee
where $\eta_{ab} = {\rm diag}\,\left[\ -\,1, 1, 1, 1\ \right]$\,,
(\,$\sqrt{-\eta} = 1$\,)\,, i.e., a {\em constant\/} Minkowskian
frame metric. The vector fields $\{\,{\bf e}_{a}\,\}$ dual to the
1--forms $\{\,\mbox{\boldmath $\omega$}^{a}\,\}$ satisfy the
relation
\be
\l{add2}
\langle\,\mbox{\boldmath $\omega$}^{a},\,{\bf e}_{b}\,\rangle
= \delta^{a}{}_{b} \ .
\ee
We choose the timelike frame vector ${\bf e}_{0}$ to be the
4-velocity of the fluid matter flow, ${\bf e}_{0} = {\bf u}$. This
setup allows for a $1+3$ split of the commutator relations as well
as of the curvature variables and their field equations (\,see
Ref. \ct{ell67}\,), part of which are constituted by the Bianchi
identities.

The commutation functions $\gamma^{a}{}_{bc}$ are defined by
\be
\l{add3}
\left[\,{\bf e}_{a}, {\bf e}_{b}\,\right] = \gamma^{c}\!_{ab}\,
{\bf e}_{c} \ ,
\ee
where the frame vectors act as differential operators, ${\bf
e}_{a}(T)$, on any geometrical objects ${\bf T}$. The purely
spatial components, $\gamma^{\alpha}{}_{\beta\gamma}$, decompose
into an object $a_{\alpha}$ and a symmetric object
$n_{\alpha\beta}$ as follows
\be
\l{add4}
\gamma^{\alpha}\!_{\beta\gamma} := 2\,a_{[\beta}\,
\delta^{\alpha}\!_{\gamma]}
+ \epsilon_{\beta\gamma\delta}\,n^{\delta\alpha} \ ,
\ee
where $\epsilon_{\alpha\beta\gamma}$ is the totally antisymmetric
3-D permutation tensor with $\epsilon_{123} = 1 = \epsilon^{123}$. 
The commutation functions with one or more indices equal to zero
can be expressed in terms of the kinematical quantities and the
quantity
\be
\l{add2a}
\Omega^{a} := {\sfrac{1}{2}}\,\eta^{abcd}\,{\bf e}_{b} \cdot
\dot{{\bf e}}_{c}\,u_{d} \ ,
\ee
(\,where $\dot{{\bf e}}_{a} := \nabla_{{\bf u}}{\bf e}_{a}$\,),
which can be interpreted as the local angular velocity of the (to
be chosen) spatial frame $\{\,{\bf e}_{\alpha}\,\}$.

Equation (\r{csildivh}) expresses the fact that one can choose a
common eigenframe for $\sig_{\alpha\beta}$ and $E_{\alpha\beta}$,
i.e., it is possible to diagonalise both tensors simultaneously.
As was shown by Barnes and Rowlingson \ct{barrow89}, it follows
from the on-diagonal components of both the $H$--constraint, Eq.
(\r{csilhconstr}), and the evolution equation for the magnetic
part, Eq. (\r{csilhdot}), that
\be
\l{osil2}
0 = n_{11} = n_{22} = n_{33} \ .
\ee
They also showed that the eigenframe $\{\,{\bf e}_{\alpha}\,\}$ of
$\sig_{\alpha\beta}$ and $E_{\alpha\beta}$ is Fermi-transported
along ${\bf u}$,
\be
\l{osil3}
\Omega^{\alpha} = 0 \ ,
\ee
a condition obtained from the vanishing off-diagonal components of
the evolution equations of both the fluid rate of shear, Eq.
(\r{csilsigdot}), and the electric part, Eq. (\r{csiledot}).

With these conditions and bearing in mind that $\om^{\mu} = 0$, one
has that $\{\,{\bf e}_{a}\,\}$ is spanned by four individually
hypersurface orthogonal basis fields, which implies that local
coordinates can be found on $\left(\,{\cal M}, \,{\bf g},\,{\bf
u}\,\right)$ with respect to which the metric tensor field ${\bf
g}$ is diagonal. This result holds for Weyl curvature tensors of
either algebraic Petrov type I or type D.

In the following analysis, we will employ tracefree--adapted
irreducible frame components for $\sig_{\alpha\beta}$ and
$E_{\alpha\beta}$, defined by (\,see, e.g., Ref. \ct{elsugg96}\,)
\bea
\begin{array}{lll}
\l{sigpm}
A_{+} :=  - \,{\sfrac{3}{2}}\,A_{11} &
= {\sfrac{3}{2}}\,(A_{22}+A_{33}) \ , &
A_{-} := {\sfrac{\sqrt{3}}{2}}\,(A_{22}-A_{33}) \ .
\end{array}
\eea
This implies
\be
\l{osil1}
A^{2} = \sfrac{1}{3}\,[\ (A_{+})^{2}  + (A_{-})^{2}\ ] \ .
\ee

Given these specialisations the following set describing silent
cosmological models according to conditions (\r{sil}) can be
derived from the general $1+3$ ONF dynamical equations
\ct{elsugg96}:\\

\noindent
The commutators:
\bea
\l{silcom1}
\left[\,{\bf e}_{0}, {\bf e}_{1}\,\right] & = & - \,{\sfrac13}\,
(\Th-2\,\sig_{+})\,{\bf e}_{1} \\
\l{silcom2}
\left[\,{\bf e}_{0}, {\bf e}_{2}\,\right] & = & - \,{\sfrac13}\,
(\Th+\sig_{+}+\sqrt{3}\,\sig_{-})\,{\bf e}_{2} \\
\l{silcom3}
\left[\,{\bf e}_{0}, {\bf e}_{3}\,\right] & = & - \,{\sfrac13}\,
(\Th+\sig_{+}-\sqrt{3}\,\sig_{-})\,{\bf e}_{3} \\
\l{silcom4}
\left[\,{\bf e}_{1}, {\bf e}_{2}\,\right] & = & - \,(a_{2}-n_{31})
\,{\bf e}_{1} + (a_{1}+n_{23})\,{\bf e}_{2} \\
\l{silcom5}
\left[\,{\bf e}_{2}, {\bf e}_{3}\,\right] & = & - \,(a_{3}-n_{12})
\,{\bf e}_{2} + (a_{2}+n_{31})\,{\bf e}_{3} \\
\l{silcom6}
\left[\,{\bf e}_{3}, {\bf e}_{1}\,\right] & = & - \,(a_{1}-n_{23})
\,{\bf e}_{3} + (a_{3}+n_{12})\,{\bf e}_{1} \ .
\eea
%

\noindent
Coupled subsystem of ordinary differential evolution equations:
\bea
\l{silthdot}
{\bf e}_{0}(\Th) & = & - \,{\sfrac{1}{3}}\,\Th^{2}
- 2\,\sig^{2} - \sfrac{1}{2}\,\mu \\
\l{silsigpdot}
{\bf e}_{0}(\sig_{+}) & = & - \,{\sfrac{1}{3}}\,(2\,\Th
-\sig_{+})\,\sig_{+} - {\sfrac{1}{3}}\,(\sig_{-})^{2} - E_{+} \\
\l{silsigmdot}
{\bf e}_{0}(\sig_{-}) & = & - \,{\sfrac{2}{3}}\,(\Th
+\sig_{+})\,\sig_{-} - E_{-} \\
{\bf e}_{0}(E_{+}) & = & - \,\sfrac{1}{2}\,\mu\,
\sig_{+} - (\Th+\sig_{+})\,E_{+} + \sig_{-}\,E_{-} \\
{\bf e}_{0}(E_{-}) & = & - \,\sfrac{1}{2}\,\mu\,
\sig_{-} - (\Th-\sig_{+})\,E_{-} + \sig_{-}\,E_{+} \\
\l{silmudot}
{\bf e}_{0}(\mu) & = & - \,\Th\,\mu \ .
\eea
Note that this set implies $\sig_{-} = 0 \Leftrightarrow
E_{-} = 0$.\\

\noindent
Remaining decoupled system of ordinary differential evolution
equations:
\bea
\l{sila1dot}
{\bf e}_{0}(a_{1}) & = & - \,{\sfrac13}\,(\Th+\sig_{+})\,
a_1 - {\sfrac{1}{\sqrt{3}}}\,\sig_{-}\,n_{23} \\
{\bf e}_{0}(a_{2}) & = & {\sfrac16}\,(-\,2\,\Th+\sig_{+}+\sqrt{3}\,
\sig_{-})\,a_2 - {\sfrac12}\,(\sig_{+}-{\sfrac{1}{\sqrt{3}}}
\,\sig_{-})\,n_{31} \\
{\bf e}_{0}(a_{3}) & = & {\sfrac16}\,(-\,2\,\Th+\sig_{+}
-\sqrt{3}\,\sig_{-})\,a_3 + {\sfrac12}\,(\sig_{+}
+{\sfrac{1}{\sqrt{3}}}\,\sig_{-})\,n_{12} \\
{\bf e}_{0}(n_{23}) & = & - \,{\sfrac13}\,(\Th+\sig_{+})\,
n_{23} - {\sfrac{1}{\sqrt{3}}}\,\sig_{-}\,a_1 \\
{\bf e}_{0}(n_{31}) & = & {\sfrac16}\,(-\,2\,\Th+\sig_{+}
+\sqrt{3}\,\sig_{-})\,n_{31} - {\sfrac12}\,(\sig_{+}
-{\sfrac{1}{\sqrt{3}}}\,\sig_{-})\,a_2 \\
\l{siln12dot}
{\bf e}_{0}(n_{12}) & = & {\sfrac16}\,(-\,2\,\Th+\sig_{+}
-\sqrt{3}\,\sig_{-})\,n_{12} + {\sfrac12}\,(\sig_{+}
+{\sfrac{1}{\sqrt{3}}}\,\sig_{-})\,a_3 \ .
\eea

\noindent
Tracefree part and trace of 3--Ricci curvature of spacelike
3--surfaces orthogonal to ${\bf u}$ (Gau\ss\ equation):
\bea
\l{sil3ric1}
{}^{*}\!S_{+} & = & - \,{\sfrac{1}{2}}\,[\ 2\,{\bf e}_{1}
(a_{1}) - {\bf e}_{2}(a_{2}) - {\bf e}_{3}(a_{3}) - 4\,(n_{23})^{2}
+ 2\,(n_{31})^{2} + 2\,(n_{12})^{2} \nonumber \\
& & - \,3\,({\bf e}_{2}-2\,a_{2})\,(n_{31})
+ 3\,({\bf e}_{3}-2\,a_{3})\,(n_{12})\ ] \nonumber \\
& = & E_{+} - {\sfrac13}\,(\Th+\sig_{+})\,
\sig_{+} + {\sfrac{1}{3}}\,(\sig_{-})^{2} \\
\l{sil3ric2}
{}^{*}\!S_{-} & = & {\sfrac{\sqrt{3}}{2}}\,[\ {\bf e}_{2}
(a_{2}) - {\bf e}_{3}(a_{3}) - 2\,(n_{31})^{2}
+ 2\,(n_{12})^{2} + 2\,({\bf e}_{1}-2\,a_{1})\,(n_{23})\nonumber \\
& &  - \,({\bf e}_{2}-2\,a_{2})\,(n_{31})
- ({\bf e}_{3}-2\,a_{3})\,(n_{12})\ ] \nonumber \\
& = & E_{-} - {\sfrac13}\,(\Th-2\,\sig_{+})\,\sig_{-}
\\ \nonumber \\
\l{silo3ric1}
0 & = & ({\bf e}_{2}-2\,a_{2})\,(n_{12}) - ({\bf e}_{3}-2\,a_{3})
\,(n_{31}) + {\bf e}_{2}(a_{3}) + {\bf e}_{3}(a_{2}) \nonumber\\
&&+ 4\,n_{31}\,n_{12} \\
\l{silo3ric2}
0 & = & ({\bf e}_{3}-2\,a_{3})\,(n_{23}) - ({\bf e}_{1}-2\,a_{1})
\,(n_{12}) + {\bf e}_{3}(a_{1}) + {\bf e}_{1}(a_{3})\nonumber\\
&&+ 4\,n_{12}\,n_{23} \\
\l{silo3ric3}
0 & = & ({\bf e}_{1}-2\,a_{1})\,(n_{31}) - ({\bf e}_{2}-2\,a_{2})
\,(n_{23}) + {\bf e}_{1}(a_{2}) + {\bf e}_{2}(a_{1})\nonumber\\
&&+ 4\,n_{23}\,n_{31} \\ \nonumber \\
\l{sil3rscl}
{}^{*}\!R & = & 2\,[\ (2\,{\bf e}_{1}-3\,a_{1})\,(a_{1})
+ (2\,{\bf e}_{2}-3\,a_{2})\,(a_{2}) + (2\,{\bf e}_{3}
-3\,a_{3})\,(a_{3}) \nonumber \\
& &  - \,(n_{23})^{2} - (n_{31})^{2} - (n_{12})^{2}
\ ] \nonumber \\
& = & 2\,\mu - \sfrac{2}{3}\,\Th^{2} + 2\,\sig^{2} \ .
\eea
Equations (\r{silo3ric1}) -- (\r{silo3ric3}) arise from the
condition that the 3--Ricci curvature tensor be diagonal.\\

\noindent
The constraint equations:
\bea
\l{sildivsig1}
0 & = & ({\bf e}_{1}-3\,a_{1})\,(\sig_{+})
- \sqrt{3}\,n_{23}\,\sig_{-} + {\bf e}_{1}(\Th) \\
\l{sildivsig2}
0 & = & ({\bf e}_{2}-3\,a_{2})\,(\sig_{+}+\sqrt{3}\,\sig_{-})
+ 3\,n_{31}\,(\sig_{+}-{\sfrac{1}{\sqrt{3}}}\,\sig_{-})
- 2\,{\bf e}_{2}(\Th) \\
\l{sildivsig3}
0 & = & ({\bf e}_{3}-3\,a_{3})\,(\sig_{+}-\sqrt{3}\,\sig_{-})
- 3\,n_{12}\,(\sig_{+}+{\sfrac{1}{\sqrt{3}}}\,\sig_{-})
- 2\,{\bf e}_{3}(\Th) \\
\l{sildive1}
0 & = & ({\bf e}_{1}-3\,a_{1})\,(E_{+}) - \sqrt{3}\,n_{23}\,E_{-}
+ \sfrac{1}{2}\,{\bf e}_{1}(\mu) \\
\l{sildive2}
0 & = & ({\bf e}_{2}-3\,a_{2})\,(E_{+}+\sqrt{3}\,E_{-})
+ 3\,n_{31}\,(E_{+} -{\sfrac{1}{\sqrt{3}}}\,E_{-})
- {\bf e}_{2}(\mu) \\
\l{sildive3}
0 & = & ({\bf e}_{3}-3\,a_{3})\,(E_{+}-\sqrt{3}\,E_{-})
- 3\,n_{12}\,(E_{+}+{\sfrac{1}{\sqrt{3}}}\,E_{-})
- {\bf e}_{3}(\mu) \\
\l{siljac1}
0 & = & ({\bf e}_{2}-2\,a_{2})\,(n_{12}) + ({\bf e}_{3}-2\,a_{3})
\,(n_{31}) + {\bf e}_{2}(a_{3}) - {\bf e}_{3}(a_{2}) \\
\l{siljac2}
0 & = & ({\bf e}_{3}-2\,a_{3})\,(n_{23}) + ({\bf e}_{1}-2\,a_{1})
\,(n_{12}) + {\bf e}_{3}(a_{1}) - {\bf e}_{1}(a_{3}) \\
\l{siljac3}
0 & = & ({\bf e}_{1}-2\,a_{1})\,(n_{31}) + ({\bf e}_{2}-2\,a_{2})
\,(n_{23}) + {\bf e}_{1}(a_{2}) - {\bf e}_{2}(a_{1}) \\
\l{silh1}
0 & = & ({\bf e}_{1}-a_{1})\,(\sig_{-}) - \sqrt{3}\,n_{23}
\,\sig_{+} \\
\l{silh2}
0 & = & ({\bf e}_{2}-a_{2})\,(\sig_{+}-{\sfrac{1}{\sqrt{3}}}
\,\sig_{-}) + n_{31}\,(\sig_{+}+\sqrt{3}\,\sig_{-}) \\
\l{silh3}
0 & = & ({\bf e}_{3}-a_{3})\,(\sig_{+}+{\sfrac{1}{\sqrt{3}}}
\,\sig_{-}) - n_{12}\,(\sig_{+}-\sqrt{3}\,\sig_{-}) \\
\l{silhdot1}
0 & = & ({\bf e}_{1}-a_{1})\,(E_{-}) - \sqrt{3}\,n_{23}\,E_{+} \\
\l{silhdot2}
0 & = & ({\bf e}_{2}-a_{2})\,(E_{+}-{\sfrac{1}{\sqrt{3}}}\,
E_{-}) + n_{31}\,(E_{+}+\sqrt{3}\,E_{-}) \\
\l{silhdot3}
0 & = & ({\bf e}_{3}-a_{3})\,(E_{+}+{\sfrac{1}{\sqrt{3}}}\,
E_{-}) - n_{12}\,(E_{+}-\sqrt{3}\,E_{-}) \ .
\eea
Here the following correspondences exist between the $1+3$ ONF form
and the $1+3$ covariant form of the constraints:
Eqs. (\r{sildivsig1}) -- (\r{sildivsig3}) correspond to Eq.
(\r{csildivsig}), Eqs. (\r{sildive1}) -- (\r{sildive3}) to Eq.
(\r{csildive}), Eqs. (\r{silh1}) -- (\r{silh3}) to
Eq. (\r{csilhconstr}), and Eqs. (\r{silhdot1}) -- (\r{silhdot3}) to
Eq. (\r{csilhdot}), respectively. Equations (\r{siljac1}) --
(\r{siljac3}) follow from the Jacobi identities.  By use of the
commutators (\r{silcom1}) -- (\r{silcom3}) and re--substitution from
known relations, one can show that the Jacobi constraints
(\r{siljac1}) -- (\r{siljac3}) as well as the Gau\ss\ constraints
(\r{sil3ric1}) -- (\r{sil3rscl}) are preserved along ${\bf u}$.

\section{Constraint analysis}
\l{ch5silcons}

In $1+3$ ONF variables, the specific silent model condition that an
irrotational dust fluid matter source induces (purely) electric
Weyl curvature of {\em zero\/} spatial rotation --- Eq.
(\r{csilhdot}) in $1+3$ covariant terms --- takes the form of Eqs.
(\r{silhdot1}) -- (\r{silhdot3}). In order for the silent
assumption, as specified by Eq. (\r{sil}), to lead to a consistent
set of dynamical equations, the zero spatial rotation condition
must be preserved along the integral curves of the preferred
timelike reference congruence ${\bf u}$. The constraints
(\r{silhdot1}) -- (\r{silhdot3}) are propagated along ${\bf u}$ by
application of the commutators (\r{silcom1}) -- (\r{silcom3}), and
it is straightforward to show that they will be preserved, given
that the conditions
\bea
\l{r1}
0 & = & E_{-}\,{\bf e}_{1}(\Th) + \sfrac{1}{2}\,\sig_{-}\,
{\bf e}_{1}(\mu) - 2\,(a_{1}\,\sig_{-}+\sqrt{3}\,n_{23}\,
\sig_{+})\,E_{+} \nonumber \\
& &   - \ 2\,(a_{1}\,\sig_{+}+\sfrac{1}{\sqrt{3}}\,
n_{23}\,\sig_{-})\,E_{-} \\
\l{r2}
0 & = & (E_{+}-\sfrac{1}{\sqrt{3}}\,E_{-})\,{\bf e}_{2}(\Th)
+ \sfrac{1}{2}\,(\sig_{+}-\sfrac{1}{\sqrt{3}}\,\sig_{-})\,
{\bf e}_{2}(\mu) \nonumber \\
& &   + \ 2\,(a_{2}-n_{31})\,(\sig_{+}+\sfrac{1}{\sqrt{3}}\,
\sig_{-})\,E_{+} \nonumber \\
& &   + \ \sfrac{2}{\sqrt{3}}\,a_{2}\,(\sig_{+}
-\sqrt{3}\,\sig_{-})\,E_{-} - \sfrac{2}{\sqrt{3}}\,n_{31}\,
(\sig_{+}+\sfrac{5}{\sqrt{3}}\,\sig_{-})\,E_{-} \\
\l{r3}
0 & = & (E_{+}+\sfrac{1}{\sqrt{3}}\,E_{-})\,{\bf e}_{3}(\Th)
+ \sfrac{1}{2}\,(\sig_{+}+\sfrac{1}{\sqrt{3}}\,\sig_{-})\,
{\bf e}_{3}(\mu) \nonumber \\
& &   + \ 2\,(a_{3}+n_{12})\,(\sig_{+}-\sfrac{1}{\sqrt{3}}\,
\sig_{-})\,E_{+} \nonumber \\
& &   - \ \sfrac{2}{\sqrt{3}}\,a_{3}\,(\sig_{+}
+\sqrt{3}\,\sig_{-})\,E_{-} - \sfrac{2}{\sqrt{3}}\,n_{12}\,
(\sig_{+}-\sfrac{5}{\sqrt{3}}\,\sig_{-})\,E_{-}
\eea
hold. The righthand sides of these equations are equivalent to the
$1+3$ covariant expression (\r{csilcurledot3}) derived above. In
the following it is assumed that the electric part of the Weyl
curvature is non--zero.

\subsection{Petrov type D}
If the spacetime geometry is spatially inhomogeneous and its Weyl
curvature tensor is of algebraic Petrov type D (\,$E_{-} = 0
\Rightarrow \sig_{-} = 0$\,), then, as Barnes and Rowlingson
\ct{barrow89} proved, it is identical to the dust models of
Szekeres \ct{sze75}, or special subcases thereof, and the equations
are consistent. This can be seen as follows. For $E_{-} = 0
\Leftrightarrow \sig_{-} = 0$, one obtains from Eq. (\r{silh1})
that $n_{23} = 0$. Then Eq. (\r{r1}) is identically satisfied,
while Eqs. (\r{r2}) and (\r{r3}) yield
\bea
\l{ptdr2}
0 & = & E_{+}\,{\bf e}_{2}(\Th) + \sfrac{1}{2}\,\sig_{+}\,
{\bf e}_{2}(\mu) + 2\,(a_{2}-n_{31})\,\sig_{+}\,E_{+} \\
\l{ptdr3}
0 & = & E_{+}\,{\bf e}_{3}(\Th) + \sfrac{1}{2}\,\sig_{+}\,
{\bf e}_{3}(\mu) + 2\,(a_{3}+n_{12})\,\sig_{+}\,E_{+} \ .
\eea
However, in that case one can derive from the constraints that
\bea
0 & = & {\bf e}_{2}(\Th) + (a_{2}-n_{31})\,\sig_{+} \nonumber \\
0 & = & {\bf e}_{3}(\Th) + (a_{3}+n_{12})\,\sig_{+} \nonumber \\
0 & = & \sfrac{1}{2}\,{\bf e}_{2}(\mu) + (a_{2}-n_{31})\,E_{+}
\nonumber \\
0 & = & \sfrac{1}{2}\,{\bf e}_{3}(\mu) + (a_{3}+n_{12})\,E_{+} \ ,
\nonumber
\eea
so that Eqs. (\r{ptdr2}) and (\r{ptdr3}) are identically
satisfied as well. A special subcase contained within the Szekeres
family are Ellis' LRS class II dust models (\,see Refs. \ct{ell67}
and \ct{elsell96}\,). Here, the further conditions
$0 = {\bf e}_{2}(f) = {\bf e}_{3}(f)$, $0 = a_{3} = n_{12}$ and
$a_{2} = n_{31}$ apply, and the equations are again consistent.

\subsection{Petrov type I}
If, on the other hand, the spacetime geometry is spatially
inhomogeneous and its Weyl curvature tensor is of algebraic Petrov
type I (\,$E_{-} \neq 0 \Rightarrow \sig_{-} \neq 0$\,), then,
contrary to what was claimed by Lesame et al \ct{lesetal95}, Eqs.
(\r{r1}) -- (\r{r3}) do {\em not\/} vanish identically, but
constitute a {\em new\/} set of constraints. Of course, they are
trivially satisfied, if a Petrov type I spacetime geometry is of
OSH Bianchi Type--I (\,${\bf e}_{\alpha}(f) = 0$, $0 = a_{\alpha} =
n_{\alpha\beta}$\,).

Equations (\r{r1}) -- (\r{r3}) can be interpreted as expressions
for the spatial 3--gradients of the fluid rate of expansion, ${\bf
e}_{\alpha}(\Th)$. Propagating them along ${\bf u}$ and
re--substituting from known relations, one obtains algebraic
expressions for the 3--gradients of the fluid's total energy
density, ${\bf e}_{\alpha}(\mu)$, in the form
\bea
\l{r4}
{\bf e}_{1}(\mu) & = & f_{1}[\ a_{1}, n_{23}, \sig_{+},
\sig_{-}, E_{+}, E_{-}, \mu\ ] \\
\l{r5}
{\bf e}_{2}(\mu) & = & g_{1}[\ a_{2}, n_{31}, \sig_{+},
\sig_{-}, E_{+}, E_{-}, \mu\ ] \\
\l{r6}
{\bf e}_{3}(\mu) & = & h_{1}[\ a_{3}, n_{12}, \sig_{+},
\sig_{-}, E_{+}, E_{-}, \mu\ ] \ .
\eea
Here, $f_{1}$, $g_{1}$ and $h_{1}$ are multivariate rational
expressions of the variables indicated. Each individual term in the
numerators therein is linear in either $a_{\alpha}$ or
$n_{\alpha\beta}$, and all terms in the expressions contain a
factor of either a power of $\sig_{-}$ or a power of $E_{-}$. As
an example, in an appendix we give the precise form of $f_{1}$
explicitly.

Propagating Eqs. (\r{r4}) -- (\r{r6}) along ${\bf u}$ and
re--substituting from known relations, one obtains algebraic
expressions for the spatial commutation functions $a_{\alpha}$ in
the form
\bea
\l{r7}
a_{1} & = & n_{23}\ f_{2}[\ \Th, \sig_{+}, \sig_{-}, E_{+},
E_{-}, \mu\ ] \\
\l{r8}
a_{2} & = & n_{31}\ g_{2}[\ \Th, \sig_{+}, \sig_{-}, E_{+},
E_{-}, \mu\ ] \\
\l{r9}
a_{3} & = & n_{12}\ h_{2}[\ \Th, \sig_{+}, \sig_{-}, E_{+},
E_{-}, \mu\ ] \ ,
\eea
where again $f_{2}$, $g_{2}$ and $h_{2}$ are multivariate rational
expressions of the variables indicated, with each individual term
containing a factor of either a power of $\sig_{-}$ or a power of
$E_{-}$.

Finally, propagating Eqs. (\r{r7}) -- (\r{r9}) along ${\bf u}$ and
re--substituting from known relations, one obtains purely algebraic
constraints of the form
\bea
\l{r10}
0 & = & n_{23}\ f_{3}[\ \Th, \sig_{+}, \sig_{-}, E_{+}, E_{-},
\mu\ ] \\
0 & = & n_{31}\ g_{3}[\ \Th, \sig_{+}, \sig_{-}, E_{+}, E_{-},
\mu\ ] \\
\l{r12}
0 & = & n_{12}\ h_{3}[\ \Th, \sig_{+}, \sig_{-}, E_{+}, E_{-},
\mu\ ] \ ,
\eea
where $f_{3}$, $g_{3}$ and $h_{3}$ are high--order multivariate
polynomial expressions of the variables indicated, with each
individual term containing a factor of either a power of $\sig_{-}$
or a power of $E_{-}$. At this stage, in $1+3$ covariant terms one
has taken the fourth spatially projected covariant time
derivative of the condition (\ref{csilhdot}) that the electric Weyl 
tensor has vanishing spatial rotation. 
We emphasise that, although the conditions (\r{r10}) --
(\r{r12}) are derived via tetrad methods, 
they reflect the covariant
property that the constraints are not identically satisfied and do
not become compatible after repeated differentiation.
It is clear from Eqs. (\r{r10}) -- (\r{r12})
that one can attempt to satisfy these conditions in four different
ways (\,modulo a cyclic permutation of the axes of the spatial
frame $\{\,{\bf e}_{\alpha}\}$\,), depending on the number of
non-zero $n_{\alpha\beta}$ variables:\\

(i) $0 = n_{23} = n_{31} = n_{12}$; this implies $a_{\alpha} = 0$,
and it is straightforward to show that this case simply corresponds
to the OSH dust models of Bianchi Type--I,

(ii) $0 = n_{31} = n_{12},$ $n_{23} \neq 0$, which implies
$0 = a_{2} = a_{3}$ and $f_{3} = 0$,

(iii) $n_{12} = 0$, $n_{23} \neq 0 \neq n_{31}$, which implies
$a_{3} = 0$ and $0 = f_{3} = g_{3}$,

(iv) all of $n_{23}$, $n_{31}$ and $n_{12}$ are non-zero, which
implies $0 = f_{3} = g_{3} = h_{3}$.   \\

Due to the particular structure of $f_{3}$, $g_{3}$ and $h_{3}$,
case (iv) could be solved, e.g., by $\sig_{-} = 0 \Leftrightarrow
E_{-} = 0$, which just reproduces the Petrov type D situation
discussed above. The issue is to see whether other non--trivial
solutions in the variables $\Th$, $\sig_{+}$, $\sig_{-}$, $E_{+}$,
$E_{-}$, and $\mu$ to the highly complex algebraic conditions $0 =
f_{3} = g_{3} = h_{3}$ could be found.  Additionally this would
involve satisfying Eqs. (\r{r4}) -- (\r{r9}), and then showing that
the time derivatives of this solution are consistent. Note that we
have {\em not\/} concluded the set of time derivatives needed to
prove the consistent result generically, rather we ceased pursuing
the consistency conditions beyond Eqs. (\r{r10}) -- (\r{r12})
because of the number of terms involved. Given that all of these
conditions were satisfied, this would establish the existence of
spatially inhomogeneous silent models with a Weyl curvature tensor
of algebraic Petrov type I.  Unfortunately, because of their
complexity, we have been unable to determine if there is such a
solution. However, we doubt that such a solution exists.
Nevertheless, {\em if\/} the contrary was true, solutions of this
type would have to arise in a quite different manner compared to
the Szekeres and OSH Type--I solutions.

Similarly, we have been unable to find non--trivial solutions in
cases (ii) and (iii). So the cases (ii) -- (iv) each require
further investigation to establish a conclusive result.

\section{Conclusion}
\l{ch5silres}

On the basis of the analysis given in the
previous section,
we conclude with the following
\begin{quotation}
{\em Conjecture:\/} 
There are no spatially
inhomogeneous irrotational dust silent models, 
whose Weyl curvature tensor is of algebraic Petrov type I.
\end{quotation}
In this case,
defined according to Eq. (\r{sil}),
the silent assumption for irrotational dust fluid
matter sources would only reproduce already known classes of
spatially inhomogeneous spacetime geometries \ct{barrow89}; they
would be very restricted. This would support the conjecture of
Matarrese et al \ct{matetal94} that realistic gravitational
collapse scenarios (\,where $\Theta < 0$\,) should involve non--zero
magnetic Weyl curvature (\,with respect to ${\bf u}$\,),
$H_{\mu\nu} \neq 0$.  This raises interesting issues about how well
the Newtonian solutions in a realistic situation can correspond to
the (more accurate) general relativistic description, because the
magnetic part of the Weyl curvature vanishes in the
Newtonian limit (\,see Ehlers and Buchert \ct{bucehl96}, confirming
the view of Ellis and Dunsby \ct{elldun96}, as opposed to claims by
Bertschinger and Hamilton \ct{ber94}\,). 
Furthermore, there is a linearisation
instability within a purely general relativistic approach to
irrotational silent cosmological models, since, as we showed in
section \r{ch5silcov}, the FLRW--linearised silent models are
consistent.

We note that analysis of the kind presented here can be conducted
either via covariant or tetrad methods. In some cases, covariant
methods succeed in the analysis of consistency (e.g. Ref. \ct{maa96}),
in some cases a
tetrad analysis seems to be required (e.g. Ref. \ct{maa97}), 
and there are
cases where either formalism can be used (e.g. Refs. \ct{ell67} 
and \ct{sze96}). We emphasise that whichever formalism is used,
this does not alter the underlying covariant nature of the problem.

Two final remarks should be made. First, the silent criterion for
barotropic perfect fluids, as introduced by Matarrese et al
\ct{matetal94,bruetal95b}, and as applied in this paper, demanded
that in mathematical terms the evolution equations within the set
of $1+3$ {\em covariant\/} dynamical equations reduce to a coupled
set of ordinary differential equations. However, it should be
pointed out that a coupled set of {\em ordinary\/} differential
equations describing evolutionary behaviour of relativistic
cosmological models can {\em also\/} be obtained from the $1+3$ ONF
dynamical equations. The best--known example of this kind are the
OSH perfect fluid models as discussed by, e.g., Ellis and MacCallum
\ct{ellmac69} (\,see also Ref. \ct{waiell96}\,). Here the reduction
of the evolution equations to a set of ordinary differential
equations is achieved by choosing a $1+3$ ONF $\{\,{\bf e}_{a}\,\}$
that is invariant under the motions induced by the $G_{3}$
isometry group of spacelike translations. Hence, according to the
ordinary differential equations criterion, the OSH perfect fluid
models can be called $1+3$ ONF silent, but, in general, {\em not\/}
$1+3$ covariantly silent. Most of the OSH perfect fluid models have
non--zero electric and magnetic parts of the Weyl curvature,
$E_{\mu\nu} \neq 0 \neq H_{\mu\nu}$, and, more importantly, {\em
non--zero\/} spatial rotation terms thereof, $I_{\mu\nu} \neq 0 \neq
J_{\mu\nu}$. However, they do not represent wave--like solutions
that convey information from one worldline to another. It is not
clear at present how to characterise this behaviour in a $1+3$
covariant manner.

Second, the demand that $H_{\mu\nu} = 0$ and $\omega^{\mu} = 0$
goes beyond what is needed to establish a silent model in the $1+3$
covariant approach to perfect fluid spacetime geometries, for all
that is required (in the spatially inhomogeneous case) is that the
pressure $p$ and the spatial rotations, $I_{\mu\nu}$ of
$E_{\mu\nu}$ and $J_{\mu\nu}$ of $H_{\mu\nu}$, should vanish. The
case (\r{sil}) was considered for simplicity (and even here, a full
solution is not yet available). At present we do not know, if there
are silent models with $E_{\mu\nu} \neq 0 \neq H_{\mu\nu}$, but $0
= I_{\mu\nu} = J_{\mu\nu}$, or if non--zero vorticity will
appreciably broaden the class of allowed silent models. Preliminary
investigations in the latter case seem to indicate that rotating
silent models are tied to severe restrictions and would therefore
be quite rare as well. However, both questions need to be
clarified.

\ack
HvE thanks the Department of Physics at Stockholm
University for a kind invitation. We thank the FRD (South Africa)
and PPARC (UK) for financial support. Throughout this work the
computer algebra package {\tt REDUCE} has been a valuable tool.\\

\noindent{\em Note added in proof:} 
It has been drawn to our attention that
the conjecture in Sec. 5 has been independently arrived at by
CF Sopuerta \cite{sop}, who gives an interesting
analysis of silent models from a different viewpoint.

\appendix
\section*{Appendix}

For illustrative purposes, in this appendix we explicitly give the
exact form of Eq. (\r{r4}). We have that
\begin{eqnarray*}
{\bf e}_{1}(\mu) & = & f_{1}[\ a_{1}, n_{23}, \sig_{+},
\sig_{-}, E_{+}, E_{-}, \mu\ ] \nonumber \\ 
& = & 4\,[\ a_{1}\,A_{1} + \sfrac{1}{2\sqrt{3}}\,n_{23}\,B_{1}
\ ]\,/\,C_{1} \ ,
\end{eqnarray*}
where
\begin{eqnarray*}
A_{1} & := & 6\,\sig_{+}\,\sig_{-}\,E_{+}\,E_{-} - (\sig_{-})^{2}
\,\mu\,E_{+} + (\sig_{-})^{2}\,(E_{+})^{2} \\ 
&& - (\sig_{-})^{2}\,
(E_{-})^{2} + 2\,E_{+}\,(E_{-})^{2} \\ 
B_{1} & := & 3\,(\sig_{+})^{2}\,\mu\,E_{-} + 24\,(\sig_{+})^{2}\,
E_{+}\,E_{-} - 6\,\sig_{+}\,\sig_{-}\,\mu\,E_{+} + 6\,\sig_{+}\,
\sig_{-}\,(E_{+})^{2} \nonumber \\
& &   - \ 6\,\sig_{+}\,\sig_{-}\,(E_{-})^{2} - (\sig_{-})^{2}
\,\mu\,E_{-} + 4\,(\sig_{-})^{2}\,E_{+}\,E_{-} \nonumber\\
& &   +\ 6\,(E_{+})^{2}
\,E_{-} + 2\,(E_{-})^{3} \\ 
C_{1} & := & 3\,\sig_{+}\,\sig_{-}\,E_{-} - (\sig_{-})^{2}\,\mu
+ (\sig_{-})^{2}\,E_{+} + 2\,(E_{-})^{2} \ .
\end{eqnarray*}
\\

The right hand sides of the remaining conditions at this and all
subsequent levels of differentiation can be obtained from a
transformation rule related to a cyclic permutation of the axes of
the spatial frame $\{\,{\bf e}_{\alpha}\,\}$. This rule is given by
making the substitutions 
\begin{eqnarray*}
& & 1\ (23) \rightarrow 2\ (31)
\rightarrow 3\ (12) \rightarrow 1\,(23) \\
&& A_{+} \rightarrow -\,
\sfrac{1}{2}\,(A_{+}+\sqrt{3}\,A_{-}) \rightarrow -\,\sfrac{1}{2}\,
(A_{+}-\sqrt{3}\,A_{-}) \rightarrow A_{+}  \\
&& A_{-} \rightarrow
\sfrac{1}{2}\,(\sqrt{3}\,A_{+}-A_{-})
\rightarrow -\,\sfrac{1}{2}\,(\sqrt{3}\,A_{+}+A_{-}) \rightarrow
A_{-} \ .
\end{eqnarray*}

\section*{References}



\begin{thebibliography}{99}

\bibitem{barrow89}
Barnes A and Rowlingson R R 1989 Irrotational perfect fluids with
a purely electric Weyl tensor {\em Class. Quantum Grav.} {\bf
6} 949

\bibitem{ber94}
Bertschinger E and Hamilton A J S 1994 Lagrangian evolution of the
Weyl tensor {\em Astrophys. J.} {\bf 435} 1

\bibitem{spa96}
Bonilla M A G, Mars M, Senovilla J M M, Sopuerta C F and Vera R
1996 Comment on `Integrability conditions for irrotational dust
with a purely electric Weyl tensor'
{\em Phys. Rev.} D {\bf 54} 6565

\bibitem{bruetal95b}
Bruni M, Matarrese S and Pantano O 1995 Dynamics of silent
universes {\em Astrophys. J.} {\bf 445} 958

\bibitem{bruetal95}
Bruni M, Matarrese S and Pantano O 1995 A local view of the
observable universe 1995 {\em Phys. Rev. Lett.} {\bf 74} 1916

\bibitem{ehl61}
Ehlers J 1993 Contributions to the relativistic mechanics of 
continuous media {\em Gen. Rel. Grav.} {\bf 25} 1225
	      
\bibitem{bucehl96}
Ehlers J and Buchert T 1997 On the Newtonian limit of the Weyl
tensor {\em Astron. Astrophys.}, in press

\bibitem{ell67}
Ellis G F R 1967 Dynamics of pressure--free matter in general
relativity {\em J. Math. Phys.} {\bf 8} 1171

\bibitem{ell71}
Ellis G F R 1971 Relativistic cosmology {\em General Relativity and
Cosmology} ed R
K Sachs (New York: Academic Press)

\bibitem{ell73}
Ellis G F R 1973 Relativistic cosmology {\em Carg\`{e}se Lectures
in Physics Vol. 6} ed E Schatzman (New York: Gordon and Breach)

\bibitem{ell96}
Ellis G F R 1996 Cosmological models from a covariant viewpoint
{\em ICGC95 Proceedings} ed T
Padmanabhan et al, to appear

\bibitem{ellbru89}
Ellis G F R and Bruni M 1989 Covariant and gauge--invariant
approach to cosmological density fluctuations {\em Phys. Rev.} D
{\bf 40} 1804

\bibitem{elldun96} 
Ellis G F R and Dunsby P K S 1997 Newtonian evolution of the Weyl 
tensor {\em Astrophys. J.} in press

\bibitem{ellmac69}
Ellis G F R and MacCallum M A H 1969 A class of homogeneous
cosmological models {\em Commun. Math. Phys.} {\bf 12} 108

\bibitem{hve96}
van Elst H 1996 Extensions and applications of $1+3$
decomposition methods in general relativistic cosmological
modelling {\em PhD thesis} University of London

\bibitem{elsell96}
van Elst H and Ellis G F R 1996 The covariant approach to LRS
perfect fluid spacetime geometries {\em Class. Quantum Grav.} {\bf
13} 1099
	
\bibitem{elsugg96}
van Elst H and Uggla C 1996 General relativistic $1+3$
orthonormal frame approach {\em Preprint}

\bibitem{goowai82}
Goode S W and Wainwright J 1982 Singularities and evolution of the
Szekeres cosmological models {\em Phys. Rev.}  D {\bf 26} 3315

\bibitem{ksmh80}
Kramer D, Stephani H, MacCallum M A H and Herlt E 1980 {\em Exact
Solutions of Einstein's Field Equations} (Cambridge: Cambridge 
University Press)

\bibitem{lesetal95}
Lesame W M, Dunsby P K S and Ellis G F R 1995 Integrability
conditions for irrotational dust with a purely electric Weyl
tensor: A tetrad analysis {\em Phys. Rev.} D {\bf 52} 3406

\bibitem{maa96}
Maartens R 1997 Linearisation instability of gravity waves?
{\em Phys. Rev.} D {\bf 55} 463

\bibitem{maaetal96}
Maartens R, Ellis G F R and Siklos S T C 1997 Local freedom
in the gravitational field {\em Class. Quantum Grav.}, in press

\bibitem{maa97}
Maartens R, Lesame W M and Ellis G F R 1997 Consistency of dust
solutions with div $H=0$ {\em Phys. Rev.} D, in press

\bibitem{matetal94}
Matarrese S, Pantano O and Saez D 1994 General relativistic
dynamics of irrotational dust: Cosmological implications {\em
Phys. Rev. Lett.} {\bf 72} 320

\bibitem{sop}
Sopuerta C F 1996 Applications of timelike and null congruences to
the construction of cosmological and astrophysical models {\em PhD
thesis} Universitat de Barcelona

\bibitem{sze75}
Szekeres P 1975 A class of inhomogeneous cosmological models {\em
Commun. Math. Phys.} {\bf 41} 56

\bibitem{sze96}
Szekeres P 1996 A tetrad--free proof of Ellis' theorem on shear--free 
dust {\em Preprint}

\bibitem{waiell96}
Wainwright J and Ellis G F R (eds) 1996 {\em Dynamical Systems in
Cosmology} (Cambridge: Cambridge University Press)

\end{thebibliography}
\end{document}